\documentclass[fleqn,usenatbib]{mnras}

\usepackage{newtxtext,newtxmath}

\usepackage[T1]{fontenc}

\DeclareRobustCommand{\VAN}[3]{#2}
\let\VANthebibliography\thebibliography
\def\thebibliography{\DeclareRobustCommand{\VAN}[3]{##3}\VANthebibliography}

\usepackage{graphicx}	
\usepackage{amsmath}	
\usepackage{newtxmath}  

\title[Repeated partial tidal disruption flares from a quiescent galaxy]{The rebrightening of a \textit{ROSAT}-selected tidal disruption event: repeated weak partial disruption flares from a quiescent galaxy?}

\author[Adam Malyali et al.]{A. Malyali$^{1}$\thanks{E-mail: amalyali@mpe.mpg.de}, Z. Liu$^{1}$, A. Rau$^{1}$,  I. Grotova$^{1}$, A. Merloni$^{1}$, A.~J.~Goodwin$^{2}$, G.~E.~Anderson$^{2}$, \newauthor J.~C.~A.~Miller-Jones$^{2}$, 
A.~Kawka$^{2}$, R.~Arcodia$^{1}$, J.~Buchner$^{1}$, K. Nandra$^{1}$, D. Homan$^{3}$, M. Krumpe$^{3}$
\\
$^{1}$Max-Planck-Institut f\"ur extraterrestrische Physik,  Giessenbachstrasse 1, 85748 Garching, Germany\\
$^{2}$International Centre for Radio Astronomy Research, Curtin University, GPO Box U1987, Perth, WA 6845, Australia\\
$^{3}$Leibniz-Institut für Astrophysik Potsdam, An der Sternwarte 16, 14482 Potsdam, Germany\\
}

\date{Accepted XXX. Received YYY; in original form ZZZ}

\pubyear{2015}

\begin{document}
\label{firstpage}
\pagerange{\pageref{firstpage}--\pageref{lastpage}}
\maketitle

\begin{abstract} 
The \textit{ROSAT}-selected tidal disruption event (TDE) candidate RX J133157.6-324319.7 (J1331), was detected in 1993 as a bright (0.2--2 keV flux of $(1.0 \pm 0.1) \times 10^{-12}$~erg~s$^{-1}$~cm$^{-2}$), ultra-soft ($kT=0.11 \pm 0.03$~keV) X-ray flare from a quiescent galaxy ($z=0.05189$). During its fifth All-Sky survey (eRASS5) in 2022, \textit{SRG}/eROSITA detected the repeated flaring of J1331, where it had rebrightened to an observed 0.2--2 keV flux of $(6.0 \pm 0.7) \times 10^{-13}$~erg~s$^{-1}$~cm$^{-2}$, with spectral properties ($kT=0.115 \pm 0.007$~keV) consistent with the \textit{ROSAT}-observed flare $\sim$30 years earlier. In this work, we report on X-ray, UV, optical, and radio observations of this system.  During a pointed \textit{XMM} observation $\sim$17 days after the eRASS5 detection, J1331 was not detected in the 0.2--2~keV band, constraining the 0.2--2~keV flux to have decayed by a factor of $\gtrsim$40 over this period. Given the extremely low probability ($\sim5\times 10^{-6}$) of observing two independent full TDEs from the same galaxy over a 30 year period, we consider the variability seen in J1331 to be likely caused by two partial TDEs involving a star on an elliptical orbit around a black hole. 
J1331-like flares show faster rise and decay timescales ($\mathcal{O}(\mathrm{days})$) compared to standard TDE candidates, with neglible ongoing accretion at late times post-disruption between outbursts.
\end{abstract}




\section{Introduction}
Benefitting from the latest generation of time-domain surveys, the past decade has seen a vast growth in the diversity of observed transients originating from galactic nuclei. These events can be crudely divided into, and described as, either `one-off’ or `repeating’ events, depending on the observed evolution of their lightcurves.

`One-off’ events, characterised by a single epoch of major transient behaviour over an observed monitoring campaign, comprise the majority of newly reported nuclear transients. These include systems where the variability is likely linked to changes in the accretion process onto a supermassive black hole, such as has been reported in previously known AGN (e.g. changing-state AGN; \citealt{frederick_new_2019,trakhtenbrot_1es_2019,ricci_destruction_2020,ricci_450_2021,frederick_family_2021}; short-rise, slowed-decay Bowen accretion flares, \citealt{trakhtenbrot_new_2019}), or due to stellar tidal disruption events (TDEs) in quiescent galaxies\footnote{Strong TDE candidates have also been reported in galaxies showing signs of previous AGN activity (e.g.~ \citealt{merloni_tidal_2015,blanchard_ps16dtm_2017,liu_tidal_2020}).} (see \citealt{saxton_x-ray_2020,van_velzen_optical-ultraviolet_2020,van_velzen_reverberation_2021,alexander_radio_2020} for recent reviews of X-ray, optical, infrared and radio observations of TDEs, respectively). Other transients, which may occur so close to the centres of galaxies that they are astrometically indistinguishable from SMBH accretion, have also been reported (e.g. supernovae exploding in the narrow-line region of AGN, \citealt{drake_discovery_2011}), or predicted to exist (e.g. stellar collisions in nuclear star clusters; \citealt{dale_red_2009}). 

Even more recently, the population of known `repeating' events has expanded. Several TDE candidates have now shown multiple major outbursts, either through their strong, double-peaked optical lightcurves (AT~2019avd, \citealt{malyali_at_2021,chen_at2019_2022}), repeated X-ray outbursts (IC~3599, \citealt{grupe_x-ray_1995,grupe_x-ray_2001,grupe_ic_2015,campana_multiple_2015}; eRASSt~J045650.3-203750, \citealt{liu_deciphering_2022}; AT~2018fyk, \citealt{wevers_rebrightening_2022}), or quasi-periodic optical outbursts potentially associated with repeated partial TDEs (ASASSN-14ko, \citealt{payne_asassn-14ko_2021}). Towards the more extreme end of known repeating transients lie the recently-discovered class of quasi-periodic eruptions (QPEs; \citealt{miniutti_nine-hour_2019,giustini_x-ray_2020,arcodia_x-ray_2021,arcodia_complex_2022}), which show large amplitude, ultra-soft X-ray outbursts, with flare duration of the order of hours, and which recur over timescales of hours to days. 

In this work, we report on the \textit{SRG}/eROSITA \citep{sunyaev_srg_2021,predehl_erosita_2021} detection of the repeated flaring of a previously reported, \textit{ROSAT}-selected TDE candidate, RXJ133157.6-324319.7 \citep{reiprich_discovery_2001,hampel_new_2022}, originating from a quiescent galaxy at $z=0.05189$ \citep{moretti_omegawings_2017}. 
In Section~\ref{sec:discovery}, we report on the detection of this system with eROSITA and follow-up observations performed with \textit{NICER} (Section~\ref{sec:nicer}), \textit{XMM} (Section~\ref{sec:xmm_analysis}), and \textit{Swift} XRT (Section~\ref{sec:swift_analysis}), as well as archival X-ray observations (Section~\ref{sec:archival_observations}), UV, optical and mid-infrared photometry (Section~\ref{sec:photometry}) and radio observations (Section~\ref{sec:radio}). We discuss the nature of the system in Section~\ref{sec:discussion}, before providing a summary in Section~\ref{sec:summary}.

All magnitudes are reported in the AB system and corrected for Galactic extinction using $A_{\mathrm{V}}=0.142$\,mag, obtained from \citep{schlafly_measuring_2011}, $R_{\mathrm{V}}=3.1$ and a Cardelli extinction law \citep{cardelli_relationship_1989}, unless otherwise stated. The effective wavelength for each filter was retrieved from the SVO Filter Profile Service\footnote{\url{http://svo2.cab.inta-csic.es/theory/fps/}}. All dates/times will be reported in universal time (UT). 

\section{Re-discovery and follow-up}\label{sec:discovery}
eRASSt~J133157.9-324321 (herein J1331) was detected on 2022-01-20 as a bright new X-ray point source in a systematic search for TDE candidates during the fifth eROSITA All-Sky survey (eRASS5). The \textit{eROSITA Science Analysis Software} (eSASS; \citealt{brunner_erosita_2022}) inferred source position was (RA$_\mathrm{J2000}$, Dec$_\mathrm{J2000})$=(13h31m57.9s, -32$^{\circ}$43$^\prime$21.2$^{\prime\prime}$), 
with a 1$\sigma$ positional uncertainty of 1.6$^{\prime\prime}$. No X-ray point source was detected within 60" of this position in each of the previous four eRASS. The eROSITA source position is consistent with a quiescent host galaxy at $z=0.05189$, with total stellar mass, $\log (M_{\star} / M_{\odot})=10.15 \pm 0.09$, and an inferred black hole mass, $\log (M_{\mathrm{BH}} / M_{\odot})= 6.5 \pm 0.2$ (appendix~\ref{sec:host_galaxy_properties}). The quiescent nature of the host is suggested by both the optical spectrum of its host galaxy (appendix~\ref{sec:optical_spectroscopy}; see also \citealt{hampel_new_2022}) and its \textit{AllWISE} \citep{wright_wide-field_2010,mainzer_initial_2014} mid-infrared colour, W1-W2=$0.05\pm0.05$~mag, far below the threshold of $\gtrsim$0.7 for mid-infrared AGN selection \citep{stern_mid-infrared_2012,assef_wise_2018}. After selecting J1331 as a promising TDE candidate, it was also realised that the host galaxy of J1331 was the same as that identified for the \textit{ROSAT}-selected TDE candidate, RXJ133157.6324319.7, first detected in outburst in 1993, and recently presented in \citet{hampel_new_2022}, with the finder chart for these transients presented in Fig.~\ref{fig:finder_chart}. The eRASS5 detection of J1331 thus suggested the remarkable rebrightening of a previously known TDE candidate, $\sim$29 years after the outburst detected by \textit{ROSAT}.

\begin{figure*}
    \centering
    \includegraphics[scale=0.9]{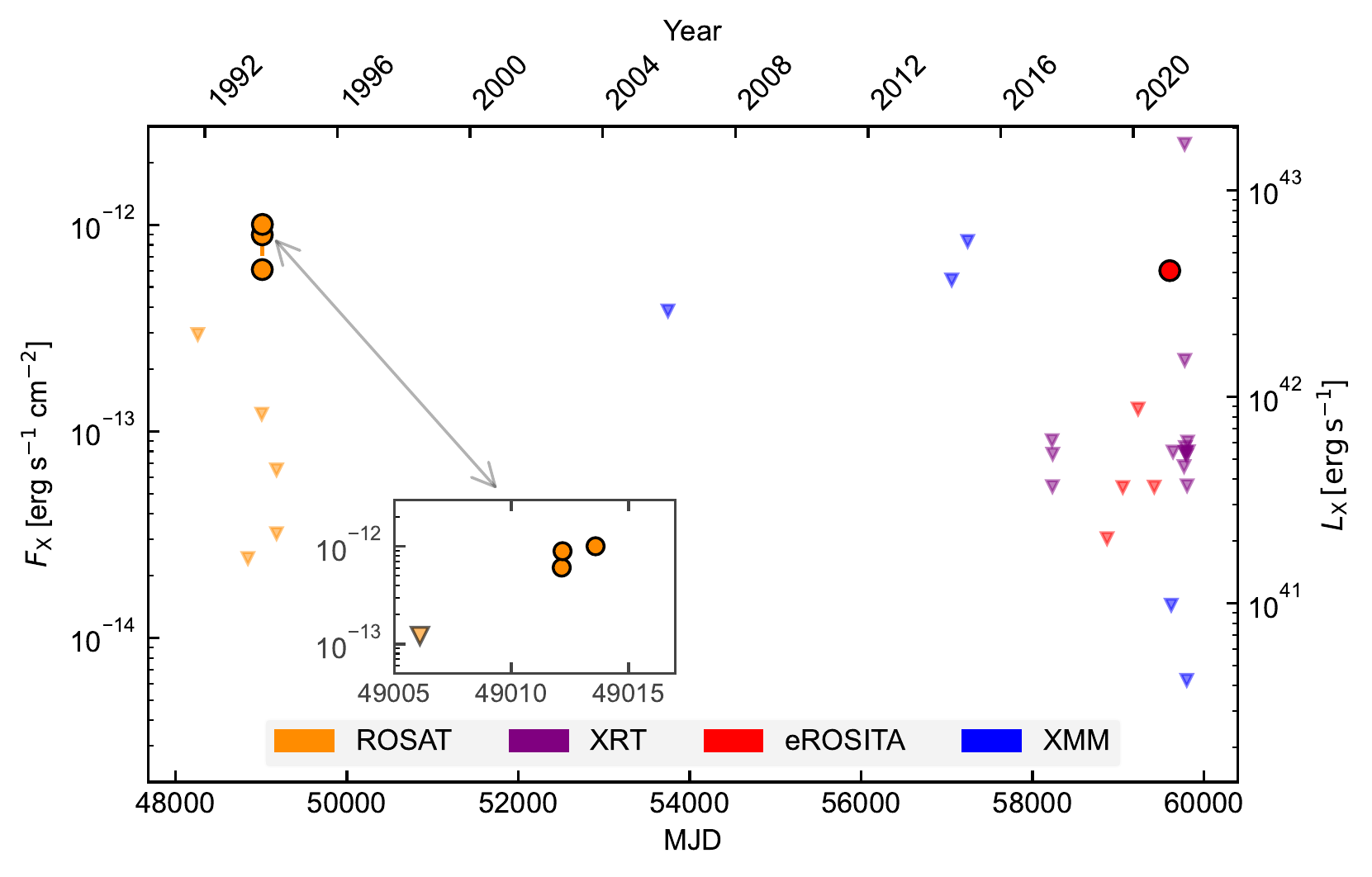}
    \caption{Long-term 0.2--2~keV lightcurve of J1331, with circular and triangle markers representing observed fluxes and 2$\sigma$ upper limits, respectively. The initial outburst was detected by \textit{ROSAT} in 1993, before being observed by eROSITA in 2022 to have rebrightened to a similar 0.2--2~keV observed flux. The X-ray spectra remained ultra-soft in each observation where the source was detected. For plotting clarity, we include the time-averaged flux measurement for eRASS5, and omit the \textit{NICER} upper limits.}
    \label{fig:long_term_lightcurve}
\end{figure*}

\subsection{eROSITA}\label{sec:erosita}
Using the eSASS task \texttt{SRCTOOL} (eSASSusers\_211214; \citealt{brunner_erosita_2022}), source (and background) spectra and lightcurves were extracted from a 60" radius source region centred on the eRASS5 inferred position, with background counts extracted from a circular annulus with inner and outer radii of 140" and 240", respectively.

eROSITA scanned the position of J1331 eight times during eRASS5, with each scan separated by $\sim$4 hours, thus spanning a $\sim$28 hour window in total. During this time, J1331 was observed to be persistently bright (Fig.~\ref{fig:erass5_lightcurve}), as opposed to showing a short-lived flaring, and was clearly detected above background in each observation. 

The eRASS5 X-ray spectra were then fitted using the Bayesian X-ray Analysis software (BXA; \citealt{buchner_x-ray_2014}), which connects the nested sampling algorithm UltraNest \citep{buchner_ultranest_2021} with the fitting environment XSPEC \citep{arnaud_xspec_1996}. The source and background spectra were jointly fit with a source plus background model, with the latter using the Principal Component Analysis (PCA) background modelling first described in \citet{simmonds_xz_2018}, and as also applied to AT~2019avd in \citet{malyali_at_2021}. The eRASS5 spectrum is well fitted by a \texttt{tbabs*zbbody} model (Fig.~\ref{fig:bxa_fit_erass5}), with the Galactic equivalent neutral hydrogen column density, $N_{\mathrm{H}}$, fixed to $3.84\times 10^{20}$\,cm$^{-2}$, the value along the line of sight to J1331 in \citet{hi4pi_collaboration_hi4pi_2016}, and  $kT=0.115 ^{+0.007}_{-0.007}$~keV. A fit with a power-law (\texttt{tbabs*zpowerlaw}) leaves large residuals between the observed data and model above 1~keV. When using the best fitting \texttt{tbabs*zbbody} model described above, the eRASS5 observed (unabsorbed) 0.2--2~keV flux for J1331 is $(6.0 \pm 0.7) \times 10^{-13}$~erg~s$^{-1}$~cm$^{-2}$ ($(8 \pm 1) \times 10^{-13}$ ~erg~s$^{-1}$~cm$^{-2}$), translating to an unabsorbed 0.2--2~keV luminosity of $(5.5 \pm 0.7) \times 10^{42}$~erg~s$^{-1}$.

J1331 was not detected in eRASS1--4, with  2$\sigma$ upper limits on the 0.2--2~keV count rate of 0.016, 0.03, 0.07 and 0.03 cts~s${^{-1}}$ in each successive eRASS (see Table~\ref{tab:x_ray_lightcurve_table} for a full log of the X-ray observations of J1331). These count rate upper limits were then converted to 0.2--2~keV flux upper limits using the best fitting spectral parameters to the eRASS5 spectrum described above.

\subsection{\textit{NICER} XTI}\label{sec:nicer}
Follow-up observations of J1331 were obtained with the X-ray Timing Instrument (XTI) on board the \textit{Neutron Star Interior Composition Explorer} observatory (\textit{NICER}; \citealt{den_herder_neutron_2016}) through pre-approved ToOs (PI: Z. Liu). \textit{NICER} observations commenced $\sim$4 days after the last eRASS5 observation, and continued for the next 15 days on a near daily basis (Table~\ref{tab:x_ray_lightcurve_table}). We first generated cleaned and screened event files using the \texttt{nicerl2} task (with default recommended parameters), before using \texttt{nibackgen3C50} \citep{remillard_empirical_2022} to generate total and background spectra for each observation ID (GTIs were filtered out using \texttt{hbgcut=0.05} and \texttt{s0cut=2}, as recommended in \citealt{remillard_empirical_2022}). ARF and RMF files were subsequently generated using the tasks \texttt{nicerarf} and \texttt{nicerrmf}, and the X-ray spectra were binned using the \citet{kaastra_optimal_2016} method to a minimum of 20 counts per bin. The total and background count rates were then estimated in the 0.4--2~keV band\footnote{The 0.4~keV lower bound here was chosen to reduce contamination from any incompletely modelled optical loading.}. J1331 is not detected at 2 sigma above background in each OBSID (Fig.~\ref{fig:nicer_rate_lc}), with 2$\sigma$ upper limits on the source count rates, inferred using $CR_{\mathrm{tot}}+2\sigma$, with $CR_{\mathrm{tot}}$ the total measured count rate, and $\sigma$ the estimated error on $CR_{\mathrm{tot}}$. The 0.4--2~keV count rates were converted to 0.2--2~keV fluxes (Table~\ref{tab:x_ray_lightcurve_table}) assuming the eRASS5 spectral model (section~\ref{sec:erosita}). \textit{NICER} observations rule out a further brightening beyond eRASS5, or a persistently bright source that rapidly `cuts-off' in brightness by the time of the \textit{XMM} observation (section~\ref{sec:xmm_analysis}).

\subsection{XMM}\label{sec:xmm_analysis}
J1331 was later observed by \textit{XMM} (P.I. Z. Liu) on 2022-02-07 (denoted XMM1), $\sim$16 days after the last eRASS5 observation, and also on 2022-08-06 (denoted XMM2). Observations were carried out with the medium filter on PN, MOS1 and MOS2. The \textit{XMM} data were reduced using HEASOFT v6.29, SAS version 20211130\_0941, and the latest calibration data files (CALDB v20210915). Following standard \textit{XMM} data reduction procedures, calibrated event files were first generated from the Observation Data Files (ODF) using the SAS tasks \texttt{emproc} and \texttt{epproc} for the MOS and PN cameras respectively. Then, periods of high background flaring were filtered out\footnote{\url{https://www.cosmos.esa.int/web/xmm-newton/sas-thread-epic-filterbackground}}. For XMM1 (XMM2), this resulted in only 4.1ks (25.7~ks), 12.8~ks (30.7~ks) and 11.8~ks (30.2~ks) of usable exposure time for PN, MOS1 and MOS2, respectively. In the subsequent analysis, only events with \texttt{PATTERN}<=4 and \texttt{FLAG}==0 were extracted for PN, whilst \texttt{PATTERN}<=12 and \texttt{FLAG}==0 filtering was applied for MOS1 and MOS2.

For XMM1, no source is detected within 30" of the host galaxy position in PN and MOS1 with detection likelihood, DETML, above 3, when running the standard \textit{XMM} source detection pipeline in the 0.2--2~keV band on the PN, MOS1, and MOS2 images. However, a source was detected in MOS2 at (RA$_\mathrm{J2000}$, Dec$_\mathrm{J2000})$=(13h31m58s, -32$^{\circ}$43$^\prime$19$^{\prime\prime}$), with a 1$\sigma$ positional uncertainty of 2$^{\prime \prime}$, 
consistent with the \textit{ROSAT} and eROSITA positions (Fig.~\ref{fig:finder_chart}). The DETML for this source is low (10.3), and the estimated observed 0.2--2~keV flux in the \texttt{emldetect} output is $(8 \pm 3) \times 10^{-15}$~erg~s$^{-1}$~cm$^{-2}$, $\sim$75$\times$ fainter than the eRASS5 observed flux. 

Given the uncertain detection of the system across all three EPIC cameras, we computed a 2$\sigma$ upper limit on the 0.2--2~keV count rate using the SAS task \texttt{eupper}. This was done using the 0.2--2~keV band images, exposure and background maps for each camera, and a 30" radius circular extraction region for the source counts (centred on the \textit{Gaia} position of the host galaxy). For XMM1, this yielded upper limits of 0.006~ct~s$^{-1}$, 0.0014~ct~s$^{-1}$ and 0.002~ct~s$^{-1}$ for PN, MOS1 and MOS2, respectively. We conservatively estimate the upper limit for the \textit{XMM} observation to that inferred from the MOS2 data, which corresponds to a 0.2--2~keV observed (unabsorbed) flux of $1\times 10^{-14}$~erg~s$^{-1}$~cm$^{-2}$ ($2\times 10^{-14}$~erg~s$^{-1}$~cm$^{-2}$), assuming the spectral model inferred from the eRASS5 observation. 
The same procedure was repeated for XMM2, where we inferred upper limits of 0.003~ct~s$^{-1}$, 0.0014~ct~s$^{-1}$ and 0.0010~ct~s$^{-1}$ for PN, MOS1 and MOS2, respectively, translating to 2$\sigma$ upper limits on the observed (unobserved) flux of $6\times 10^{-15}$ ~erg~s$^{-1}$~cm$^{-2}$ ($1\times10^{14}$ ~erg~s$^{-1}$~cm$^{-2}$).

\subsection{\textit{Swift} XRT}\label{sec:swift_analysis}
Additional \textit{Swift} XRT \citep{burrows_swift_2005} observations of J1331 were performed between 2022-02-27 and 2022-08-24\footnote{The delay between the eRASS5 and \textit{Swift} observations stemmed from the January 2022 reaction wheel failure on-board the \textit{Swift} observatory.}. The XRT observations were performed in photon counting mode, with the data analysed using the UK Swift Science Data Centre's (UKSSDC) online XRT product building tool \citep{evans_online_2007,evans_methods_2009}. No source was detected in the 0.3--2~keV band at the position of J1331 in any follow-up observation.
The 0.3--2~keV count rates were converted to 0.2--2~keV fluxes using webPIMMs\footnote{\url{https://heasarc.gsfc.nasa.gov/cgi-bin/Tools/w3pimms/w3pimms.pl}}, assuming the same spectral model as from the eROSITA eRASS5 detection, with the fluxes presented in Table~\ref{tab:x_ray_lightcurve_table}.

\subsection{Archival X-ray observations}\label{sec:archival_observations}
A detailed analysis of the ultra-soft outburst from RXJ133157.6324319.7, detected by pointed \textit{ROSAT} PSPC observations in the early 1990s, was previously performed in \citet{hampel_new_2022}. In summary, the flare was characterised by an 8x increase in the 0.1--2.4~keV flux, relative to a 2$\sigma$ upper limit, over an 8 day period (and a net increase in the same band by a factor of at least 40 relative to the deepest upper limit available). The X-ray spectrum at peak observed brightness was well fitted by a blackbody with $kT=0.11\pm 0.03$~keV. The system was then not detected in two PSPC observations $\sim$165 days later, where it had faded by a factor of at least 30 relative to the peak observed \textit{ROSAT} flux.

To construct a long-term 0.2--2~keV lightcurve, the 0.1--2.4~keV \textit{ROSAT} PSPC lightcurve data in Table~1 of \citet{hampel_new_2022} was converted into 0.2--2~keV band fluxes using webPIMMS, assuming the best fitting spectral model to the \textit{ROSAT} spectrum found in \citet{hampel_new_2022}. Then, the 2$\sigma$ upper limits from \textit{ROSAT} Survey, \textit{XMM} Slew and Swift XRT observations were computed using the \textit{High-Energy Lightcurve Generator} server (HILIGT; \citealt{saxton_hiligt_2021,konig_hiligt_2021}); the archival fluxes are presented in Fig.~\ref{fig:long_term_lightcurve} and Table~\ref{tab:x_ray_lightcurve_table}.

\subsection{UV, optical and mid-infrared photometry}\label{sec:photometry}
J1331 was observed both before (Section~\ref{sec:archival_observations}) and after (Section~\ref{sec:swift_analysis}) the eRASS5-detected outburst by \textit{Swift} XRT and UVOT (UVM2 filter; \citealt{roming_swift_2005}). To search for transient UV emission, aperture photometry was performed on the level 2 UVOT sky images (downloaded from the UKSSDC) using the \texttt{uvotsource} task (HEASOFT v6.29, CALDB v20201215). Source counts were extracted from a circular aperture of 5$^{\prime \prime}$ radius, centred on the \textit{Gaia} position of the host of J1331, and background counts were extracted from a source-free region of radius 15$^{\prime \prime}$. The measured UVM2 magnitudes in the follow-up observations 
are consistent with the archival measured UVM2 magnitudes on the 2018-04-18, 2018-04-22, 2018-04-26 (Table~\ref{tab:uvot_photometry}).
 
No significant optical variability is seen in the $\sim$6 years before the eRASS5 outburst (57500$\lesssim \mathrm{MJD} \lesssim$59500) in the forced photometry lightcurve provided by ATLAS \citep{tonry_atlas_2018} (Fig.~\ref{fig:atlas_neowise_lightcurve}). Lastly, we note that no major variability is detected above the host galaxy emission within the \textit{NEOWISE} mid-infrared lightcurve between MJD$\sim$56680 and 59400 (Fig.~\ref{fig:atlas_neowise_lightcurve}), which was generated using the procedure described in section~3.2 of \citet{malyali_at_2021}. 

\subsection{Radio}\label{sec:radio}
We observed the coordinates of J1331 on 2022 Mar 02 with the Australia Telescope Compact Array (ATCA) radio telescope in 6\,km configuration, using the 4cm dual receiver with central frequencies 5.5/9\,GHz, each with a 2~GHz bandwidth split into 2049$\times$1~MHz spectral channels, and for a total of 150\,min on source. Data were reduced following standard procedures in the Common Astronomy Software Applications \citep{mcmullin_casa_2007,casa-team_casa_2022}. We used 1934-638 for flux and bandpass calibration and 1336-260 for phase calibration. Images of the target field were created using the CASA task \texttt{tclean}. No source was detected at the location of J1331 at either frequency band with a 3$\sigma$ upper limit of 73.5$\mu$Jy/bm at 5.5\,GHz and 54$\mu$Jy/bm at 9\,GHz. Additionally, no source was detected in a stacked 5.5 and 9\,GHz image, with a 3$\sigma$ upper limit of 57.9$\mu$Jy/bm at a central frequency of 7.3\,GHz. 

\section{Discussion}\label{sec:discussion}
Comparing the X-ray lightcurve of J1331 with other ultra-soft nuclear transients (Fig.~\ref{fig:nuclear_transient_comparison}) from galaxies that were recently quiescent, or hosted low luminosity AGN, then J1331 decays faster than the majority of other X-ray bright TDEs\footnote{Ignoring short timescale flaring behaviour seen in some TDE candidates, such as AT~2019ehz \citep{van_velzen_seventeen_2021}.}, but decays over much longer timescales than the bursts typically seen in QPEs (burst durations $\lesssim$30~ks, or $\lesssim$0.3~days; \citealt{{miniutti_nine-hour_2019,giustini_x-ray_2020,arcodia_x-ray_2021,arcodia_complex_2022}}). 

Given the quiescent nature of the host galaxy, and the ultra-soft X-ray spectrum, an AGN origin for J1331 is disfavoured. We also rule out a mechanism similar to that producing the X-ray flares observed in Sgr~A* (e.g. \citealt{neilsen_chandra_2013,ponti_fifteen_2015,yuan_systematic_2016,ponti_powerful_2017,mossoux_continuation_2020}), as the latter are clearly observationally distinct to J1331, with respect to the flaring timescales (Sgr~A* flare durations $\lesssim 10^4$~s; \citealt{mossoux_continuation_2020}), spectral properties (flaring X-ray emission in Sgr~A* is hard and likely synchrotron, 
~e.g. \citealt{ponti_powerful_2017}), and peak observed luminosity (bolometric luminosity of Sgr~A* is $\sim10^{36}$~erg~s$^{-1}$; \citealt{genzel_galactic_2010}). 
Arguments against a Galactic origin for this system have previously been presented in \citet{hampel_new_2022}.

Ultra-soft X-ray flares from quiescent galaxies have previously been considered as a reliable signature of a TDE (e.g.~\citealt{zabludoff_distinguishing_2021}). However, the current theoretically predicted TDE rates are $\gtrsim 10^{-4}$~yr$^{-1}$~galaxy$^{-1}$ \citep{stone_rates_2020}, so it would be exceptionally unlikely to have observed two independent tidal disruption flares occuring within the same galaxy over a $\sim$30 year timescale (Poisson probability $\sim 5\times 10^{-6}$; Fig.~\ref{fig:poisson_probability_tde_rate}); a more exotic class of TDE would need to be invoked to explain J1331. 

One such possibility, discussed in \citet{hampel_new_2022}, is that J1331 was produced by a TDE involving a supermassive black hole binary (SMBHB). This scenario was partly proposed in an attempt to explain the fast X-ray brightening observed by \textit{ROSAT}, since such TDEs may have highly non-monotonic decays of their X-ray lightcurves. This stems from the gravitational interaction between the companion BH and the debris streams, which may cause large perturbations to the orbits of the less bound debris and cause their chaotic evolution, as well as a complex evolution of the accretion rate over time. 
\citet{liu_milliparsec_2014,ricarte_tidal_2016,coughlin_tidal_2017} predict these systems to show sharp dips and rises in the X-ray lightcurve rate (of $\sim$1--2 orders of magnitude), on timescales of the order of the binary orbital period \citep{liu_milliparsec_2014,ricarte_tidal_2016}, although \citet{coughlin_tidal_2017} find highly variable accretion rates between different simulation runs and over timescales shorter than the SMBHB orbital periods (i.e. there still seems to be quite large uncertainties in the theoretically predicted lightcurves of TDEs involving SMBHBs). 

Under the SMBHB scenario, both the eROSITA and \textit{ROSAT} observations would have had to have sampled a `dipping’, or `brightening from a dip’, phase of the X-ray lightcurve, respectively. For binary orbital periods of the order of $\sim$months, assuming $\sim$mpc binary separation as in \citet{liu_milliparsec_2014}, then it would be quite fortuitous for us to have observed such behaviour. 
Furthermore, there is importantly no evidence for late time X-ray rebrightening episodes in the months after each outburst, as seen by \textit{XMM} and \textit{Swift} (Fig.~\ref{fig:long_term_lightcurve}), which one might expect to have observed given that the accretion rate is predicted to eventually revert back to the $t^{-5/3}$ decay following ‘dips’ (e.g. Fig.~12 in \citealt{coughlin_tidal_2017}). We would therefore disfavour J1331 being caused by a full TDE around a SMBHB, given the fine tuning needed in order to match observations. 

A more feasible scenario is that both outbursts were driven by a partial tidal disruption event (pTDE), potentially of the same object. Unless the pTDE rate is orders of magnitude larger than currently estimated in the literature \citep{stone_rates_2016,chen_light_2021,zhong_revisit_2022}, then both outbursts would likely be related to the same star being disrupted by the same black hole (i.e. the star should have survived the initial encounter). 
Considering that the recurrence timescale of J1331 is $\lesssim 30$~years, then it is also difficult to reconcile this with theoretical predictions for the recurrence timescales of flares in pTDEs where the star was initially scattered onto a parabolic orbit around the black hole ($\gtrsim 400$~years, e.g.~ \citealt{ryu_tidal_2020}). Instead, the flaring may have been driven by the repeated stripping of a star on an elliptical orbit by the disrupting SMBH (see \citealt{hayasaki_finite_2013} for a discussion on potential origins for such stars). 
This scenario would be further supported by both the relatively small amount of inferred energy emitted in the eROSITA-detected outburst\footnote{Assuming a similar temporal evolution for both the eROSITA-detected and \textit{ROSAT}-detected outbursts- see section~\ref{sec:outburst_inference}.} of $(5^{+6}_{-3})\times 10^{49}$~erg, 
corresponding to an accreted mass of $(5^{+7}_{-2})\times 10^{-4} ( \epsilon / 0.05)^{-1}$~ $\mathrm{M}_{\odot}$, where $\epsilon$ is the radiative efficiency of accretion, and also by the extremely low $L_{\mathrm{X}}$ at late-times (as suggested by the non-detection and deep upper limits in XMM2), since elliptical TDEs are predicted to produce short-lived, finite accretion bursts \citep{hayasaki_finite_2013}. 
Given this, and that the radio observations were taken $\sim$40 days after the eRASS5 flare (section~\ref{sec:radio}), then we note that we may have missed any associated jet or outflow launched in this event, as seen in other TDE candidates (e.g. \citealt{goodwin_at2019azh_2022}).

\begin{figure}
    \centering
    \includegraphics[scale=0.8]{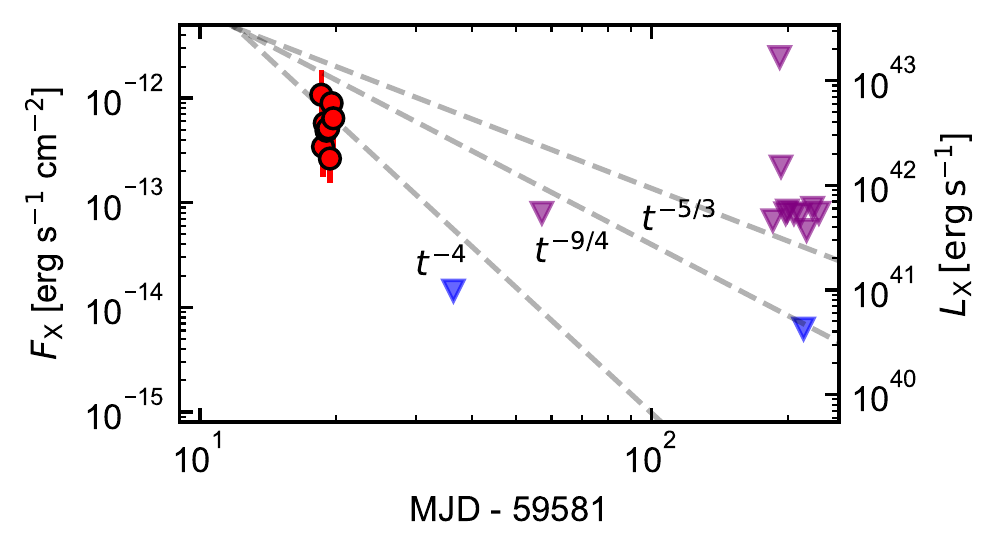}
    \caption{Zoom-in on the first eROSITA-detected outburst in 2022, along with multiple power-law decay slopes plotted in grey dashed lines. The decay slope appears to be much steeper than the canonical $t^{-5/3}$ decay predicted for TDEs with a uniform distribution of specific energies, and appears more consistent with a $t^{-4}$ decay, as predicted in \citet{ryu_tidal_2020}. We assume a peak MJD of 59593 for the X-ray outburst, and roughly estimate the MJD of disruption to be 59581 (section~\ref{sec:outburst_inference}). The markers follow the same legend as for Fig.~\ref{fig:long_term_lightcurve}.}
    \label{fig:erosita_outburst_powerlaw_decay_lines}
\end{figure}

The case for a repeated pTDE is further enhanced by the fast rise and decay timescales seen with \textit{ROSAT} and eROSITA. Compared with full disruptions, pTDEs only strip the outermost layers of the star, with the specific energy distribution of the debris, $\mathrm{d}M/\mathrm{d}E$, differing from full TDEs (e.g.~\citealt{coughlin_partial_2019,miles_fallback_2020,ryu_tidal_2020}). 
Since the mass fallback rate, $\dot{M}_{\mathrm{fb}}(t)$, scales $\propto \mathrm{d}M/\mathrm{d}E$, then $\dot{M}_{\mathrm{fb}}(t)$ is also predicted to differ between full and pTDEs. \citet{ryu_tidal_2020} find that the narrower spreads in $\mathrm{d}M/\mathrm{d}E$ for pTDEs can yield $\dot{M}_{\mathrm{fb}}(t) \propto t^{-p}$, where $p \sim 2-5$, more consistent with what is observed in J1331 (Fig.~\ref{fig:erosita_outburst_powerlaw_decay_lines}), and much steeper than a canonical $t^{-5/3}$ decline predicted for the mass fallback rate in full TDEs \citep{rees_tidal_1988,phinney_manifestations_1989}.

Lastly, although the mass fallback in weak pTDEs may evolve over shorter timescales relative to full TDEs, the viscous timescale, $t_{\mathrm{visc}}$, still needs to be shorter than the minimum orbital period of the stellar debris so that the X-ray luminosity traces the mass fallback rate (assuming a constant radiative efficiency, negligible obscuration of the soft X-rays, and negligible disc cooling). Considering $t_{\mathrm{visc}}\sim \alpha ^{-1}(H/R)^{-2}\Omega ^{-1}(r)$, where $\alpha$ is the viscosity parameter \citep{shakura_black_1973}, $H$ and $R$ the scale height and width of the disc, and $\Omega ^{-1}(r)$ the orbital period at distance $r$ from the black hole, then $t_{\mathrm{visc}}\sim 0.4 (\alpha / 0.1)^{-1}(H/R)^{-2}$~days at the circularisation radius ($\sim 2 R_{\mathrm{tidal}} / \beta$, where $R_{\mathrm{tidal}}$ and $\beta$ are the tidal radius and impact parameter for the disruption). A geometrically thick disc ($H/R \sim 1$), as may be expected to form for super-Eddington mass fallback rates, would be needed to reproduce accretion timescales of the order $\sim$days as seen in J1331. However, it is currently unclear how the stellar debris might circularise so efficiently in a weak pTDE (see \citealt{bonnerot_formation_2021} for a review on accretion flow formation in TDEs), and we also highlight here that similar concerns have recently been raised for explaining the short X-ray flare durations observed in QPEs via an accretion origin (e.g.~\citealt{krolik_quasi-periodic_2022,lu_quasi-periodic_2022}). Although future simulations would likely be needed to explore the debris circularisation in J1331-like events, alternative origins for the X-ray emission may be from compression shocks of the debris streams at pericentre (e.g.~\citealt{steinberg_origins_2022}), or circularisation shocks from debris stream collisions \citep{krolik_quasi-periodic_2022,lu_quasi-periodic_2022}. 

\section{Summary}\label{sec:summary}
J1331 is a repeating X-ray transient associated to a quiescent galaxy at $z=0.05189$, which we consider to be consistent with a scenario involving two weak pTDEs. Whilst several previously reported pTDE candidates have occurred in galaxies hosting an AGN, we highlight that the host of J1331 is quiescent. The main properties of J1331 can be summarised as follows:
\begin{enumerate}
    \item J1331 was first detected by \textit{ROSAT} in 1993 \citep{hampel_new_2022}, where it had shown an ultra-soft ($kT=0.11\pm 0.03$~keV) flaring by a factor of at least 40 relative to a previous 2$\sigma$ upper limit. The outburst also showed a fast rise, where it had brightened by a factor of eight over an 8 day period. The system was subsequently not detected in a deep pointed \textit{ROSAT} observation $\sim$165 days afterwards, as well as in \textit{XMM} Slew, and \textit{Swift} XRT observations performed between 2006 and 2018 (Table~\ref{tab:x_ray_lightcurve_table}).
    \item After not being detected by eROSITA in its first four eRASS, J1331 was observed to have brightened in eRASS5 to a 0.2--2~keV flux of $(6.0 \pm 0.7) \times 10^{-13}$~erg~s$^{-1}$~cm$^{-2}$. The eRASS5 spectrum is ultra-soft ($kT=0.115 ^{+0.007}_{-0.007}$~keV), and is consistent with the $kT$ inferred from the \textit{ROSAT}-observed flare in 1993.
    \item J1331 was not detected during pointed \textit{XMM} observations and \textit{Swift} XRT observations when followed up after the eRASS5 detection; the first (second) \textit{XMM} observation constrains the 0.2--2~keV flux to decay by a factor of $\gtrsim$40 ($\gtrsim$100) over a 17 ($\sim$200) day period after the eRASS5 observation. The faint 0.2--2~keV X-ray luminosities ($< 7\times 10^{40}$~erg~s$^{-1}$, unabsorbed) at $\sim200$~days post-peak brightness, inferred via the second \textit{XMM} observation (Table~\ref{tab:x_ray_lightcurve_table}), may be due to a late-time drop off in the mass fallback rate once the disruption episode is over.
    \item Combined with the fast rise timescale seen by \textit{ROSAT}, then J1331-like outbursts are short lived (rise and decay timescales of $6 ^{+1}_{-1}$~days and $3.9^{+0.1}_{-0.1}$~days, respectively; appendix~\ref{sec:outburst_inference}) and evolve over shorter timescales relative to full TDEs.  
    \item J1331 has only been observed to show transient emission in the 0.2--2~keV band, with no transient optical, UV, or radio emission observed in follow-up observations.
\end{enumerate}

We conclude by noting that J1331 appears to fill in the continuum of observed soft X-ray outbursts from quiescent galaxies, lying in between QPEs and TDEs with respect to its rise and decay timescales (Fig.~\ref{fig:nuclear_transient_comparison}), although the recurrence timescales are much longer than in the current sample of QPEs. Additional follow-up observations will be scheduled in order to more tightly constrain the recurrence timescales of outbursts from J1331. Future planned X-ray missions geared towards exploiting the X-ray transient sky, such as the \textit{Einstein Probe} \citep{yuan_einstein_2018}, will likely be sensitive towards detecting similar partial disruptions; for these missions, the eROSITA All-Sky survey data may play an important role by providing a long-term baseline towards which new candidates can be identified.
Given the faster decay timescales of J1331-like systems, 
then we would advocate promptly triggering high-cadence X-ray follow-up in order to better constrain the evolution of the accretion rate in future candidates. 

\section*{Acknowledgements}
AM thanks Taeho Ryu for very useful discussions whilst preparing the manuscript. AM acknowledges support by DLR under the grant 50 QR 2110 (XMM\_NuTra, PI: Z. Liu). This work was supported by the Australian government through the Australian Research Council’s Discovery Projects funding scheme (DP200102471). We would like to thank the referee for a constructive report that improved the quality of the paper.

This work is based on data from eROSITA, the soft X-ray instrument aboard SRG, a joint Russian-German science mission supported by the Russian Space Agency (Roskosmos), in the interests of the Russian Academy of Sciences represented by its Space Research Institute (IKI), and the Deutsches Zentrum für Luft- und Raumfahrt (DLR). The SRG spacecraft was built by Lavochkin Association (NPOL) and its subcontractors, and is operated by NPOL with support from the Max Planck Institute for Extraterrestrial Physics (MPE).

The development and construction of the eROSITA X-ray instrument was led by MPE, with contributions from the Dr. Karl Remeis Observatory Bamberg \& ECAP (FAU Erlangen-Nuernberg), the University of Hamburg Observatory, the Leibniz Institute for Astrophysics Potsdam (AIP), and the Institute for Astronomy and Astrophysics of the University of T{\"u}bingen, with the support of DLR and the Max Planck Society. The Argelander Institute for Astronomy of the University of Bonn and the Ludwig Maximilians Universit{\"a}t Munich also participated in the science preparation for eROSITA.

The eROSITA data shown here were processed using the eSASS software system developed by the German eROSITA consortium.

This work made use of data supplied by the UK Swift Science Data Centre at the University of Leicester.

The Australia Telescope Compact Array is part of the Australia Telescope National Facility (\url{https://ror.org/05qajvd42}) which is funded by the Australian Government for operation as a National Facility managed by CSIRO. We acknowledge the Gomeroi people as the traditional owners of the Observatory site.

The Legacy Surveys consist of three individual and complementary projects: the Dark Energy Camera Legacy Survey (DECaLS; Proposal ID 2014B-0404; PIs: David Schlegel and Arjun Dey), the Beijing-Arizona Sky Survey (BASS; NOAO Prop. ID \#2015A-0801; PIs: Zhou Xu and Xiaohui Fan), and the Mayall z-band Legacy Survey (MzLS; Prop. ID \#2016A-0453; PI: Arjun Dey). DECaLS, BASS and MzLS together include data obtained, respectively, at the Blanco telescope, Cerro Tololo Inter-American Observatory, NSF’s NOIRLab; the Bok telescope, Steward Observatory, University of Arizona; and the Mayall telescope, Kitt Peak National Observatory, NOIRLab. Pipeline processing and analyses of the data were supported by NOIRLab and the Lawrence Berkeley National Laboratory (LBNL). The Legacy Surveys project is honored to be permitted to conduct astronomical research on Iolkam Du’ag (Kitt Peak), a mountain with particular significance to the Tohono O’odham Nation.

NOIRLab is operated by the Association of Universities for Research in Astronomy (AURA) under a cooperative agreement with the National Science Foundation. LBNL is managed by the Regents of the University of California under contract to the U.S. Department of Energy.

This project used data obtained with the Dark Energy Camera (DECam), which was constructed by the Dark Energy Survey (DES) collaboration. Funding for the DES Projects has been provided by the U.S. Department of Energy, the U.S. National Science Foundation, the Ministry of Science and Education of Spain, the Science and Technology Facilities Council of the United Kingdom, the Higher Education Funding Council for England, the National Center for Supercomputing Applications at the University of Illinois at Urbana-Champaign, the Kavli Institute of Cosmological Physics at the University of Chicago, Center for Cosmology and Astro-Particle Physics at the Ohio State University, the Mitchell Institute for Fundamental Physics and Astronomy at Texas A\&M University, Financiadora de Estudos e Projetos, Fundacao Carlos Chagas Filho de Amparo, Financiadora de Estudos e Projetos, Fundacao Carlos Chagas Filho de Amparo a Pesquisa do Estado do Rio de Janeiro, Conselho Nacional de Desenvolvimento Cientifico e Tecnologico and the Ministerio da Ciencia, Tecnologia e Inovacao, the Deutsche Forschungsgemeinschaft and the Collaborating Institutions in the Dark Energy Survey. The Collaborating Institutions are Argonne National Laboratory, the University of California at Santa Cruz, the University of Cambridge, Centro de Investigaciones Energeticas, Medioambientales y Tecnologicas-Madrid, the University of Chicago, University College London, the DES-Brazil Consortium, the University of Edinburgh, the Eidgenossische Technische Hochschule (ETH) Zurich, Fermi National Accelerator Laboratory, the University of Illinois at Urbana-Champaign, the Institut de Ciencies de l’Espai (IEEC/CSIC), the Institut de Fisica d’Altes Energies, Lawrence Berkeley National Laboratory, the Ludwig Maximilians Universitat Munchen and the associated Excellence Cluster Universe, the University of Michigan, NSF’s NOIRLab, the University of Nottingham, the Ohio State University, the University of Pennsylvania, the University of Portsmouth, SLAC National Accelerator Laboratory, Stanford University, the University of Sussex, and Texas A\&M University.

BASS is a key project of the Telescope Access Program (TAP), which has been funded by the National Astronomical Observatories of China, the Chinese Academy of Sciences (the Strategic Priority Research Program “The Emergence of Cosmological Structures” Grant \# XDB09000000), and the Special Fund for Astronomy from the Ministry of Finance. The BASS is also supported by the External Cooperation Program of Chinese Academy of Sciences (Grant \# 114A11KYSB20160057), and Chinese National Natural Science Foundation (Grant \# 12120101003, \# 11433005).

The Legacy Survey team makes use of data products from the Near-Earth Object Wide-field Infrared Survey Explorer (NEOWISE), which is a project of the Jet Propulsion Laboratory/California Institute of Technology. NEOWISE is funded by the National Aeronautics and Space Administration.

The Legacy Surveys imaging of the DESI footprint is supported by the Director, Office of Science, Office of High Energy Physics of the U.S. Department of Energy under Contract No. DE-AC02-05CH1123, by the National Energy Research Scientific Computing Center, a DOE Office of Science User Facility under the same contract; and by the U.S. National Science Foundation, Division of Astronomical Sciences under Contract No. AST-0950945 to NOAO.

M.K. acknowledges support from DFG grant KR
3338/4-1.  D.H. is supported by DLR grant FKZ
50OR2003.

\section*{Data Availability}
The eRASS1-4 data taken within the German half of the eROSITA sky is currently planned to be made public by Q2 2024, whilst the eRASS5 data is scheduled to become public by Q2 2026. The Swift data is available to download through the UK Swift Data Science website\footnote{\url{https://www.swift.ac.uk/archive/index.php}}, whilst the NICER data is accessible through NASA’s HEASARC interface\footnote{\url{https://heasarc.gsfc.nasa.gov/docs/nicer/nicer_archive.html}}. Publicly available ATLAS data can be accessed through the ATLAS forced photometry service\footnote{\url{https://fallingstar-data.com/forcedphot/}}, and \textit{NEOWISE} lightcurves can be accessed through the IRSA web portal\footnote{\url{https://irsa.ipac.caltech.edu/applications/wise/}}. ATCA data are stored in the Australia Telescope Online Archive\footnote{\url{https://atoa.atnf.csiro.au/}}, and will become publicly accessible 18 months from the date of observation. The \textit{XMM} data will become public after the propietory period expires (2023-08-30). Follow-up optical spectra will likely remain private at least until the release of the forthcoming eROSITA-selected TDE population paper, but could be made available upon reasonable request.



\bibliographystyle{mnras}
\bibliography{rosat_tde} 



\appendix
\section{Host galaxy properties}\label{sec:host_galaxy_properties}
\begin{figure}
    \centering
    \includegraphics[scale=0.95]{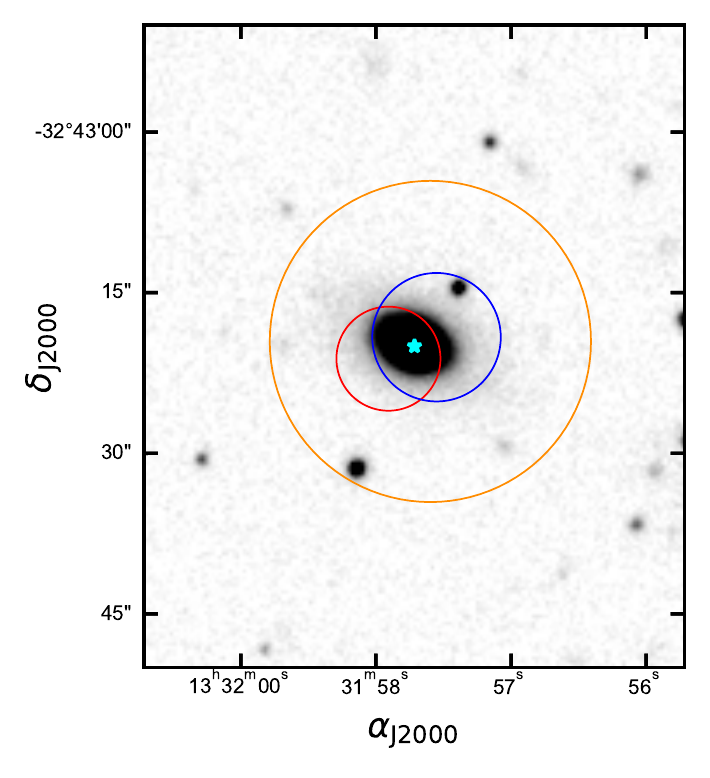}
    \caption{Legacy Survey DR10~(early) $g$-band cutout image of the sky region surrounding eRASSt~J133158-324321. The dark orange circle is the error circle for RXJ133157.6324319.7 inferred from \textit{ROSAT} pointed observations in \citet{hampel_new_2022}, whilst the red and blue circles denote the 3$\sigma$ error circles on the source position inferred from eROSITA and \textit{XMM} MOS2 observations (although the detection of J1331 in the first \textit{XMM} observation is uncertain and we quote upper limits on the count rates for this in section~\ref{sec:xmm_analysis}, we include it in this finder chart for completeness). The cyan star marks the \textit{Gaia} EDR3 \citep{gaia_collaboration_gaia_2021} position of the host galaxy.}
    \label{fig:finder_chart}
\end{figure}

Using the correlation reported in \citet{kettlety_galaxy_2018} between galaxy total stellar mass, $M_{\star}$, and luminosity in the \textit{WISE} $W1$-band, $L_{\mathrm{W1}}$,
then we infer $\log (M_{\star} / M_{\odot})=10.15 \pm 0.09$ for the host galaxy. Combining this with $M_{\mathrm{BH}}-M_{\star}$ relation in \citet{reines_relations_2015}, suggests a black hole mass of $\log (M_{\mathrm{BH}} / M_{\odot})= 6.5 \pm 0.2$. The finder chart for J1331 is presented in Fig~\ref{fig:finder_chart}. 

\section{Optical spectroscopy}\label{sec:optical_spectroscopy}

\textit{LCO spectrum (2022-02-12)}: J1331 was observed with the low dispersion FLOYDS spectrograph on the LCOGT 2m telescope at Siding Spring Observatory operated by the Las Cumbres Observatory (LCO; \citealt{brown_cumbres_2013}) on 2022 February 12 (proposal ID CON2022A-001, PI: M. Salvato). We obtained an exposure of 1800 seconds using the “red/blu” grism and the 2” slit oriented along the parallactic angle. The spectrum has a wavelength range of 3200-10000A with dispersions of 3.51A/pixel and 1.74 A/pixel in the  blue (3200-5700A) and red (5400-10000A) bands, respectively. The data were reduced and calibrated using the automatic FLOYDS pipeline. The HgAr and Zn lamps were used for wavelength calibration and a Tungsten-Halogen + Xenon lamp for flat fielding. A sensitivity function from the FLOYDS archive was used for flux calibration.

\textit{WiFeS spectrum (2022-05-09)}: We observed J1331 with the Wide Field Spectrograph (WiFeS; \citealt{dopita_wide_2010}) on the ANU 2.3m telescope at Siding Spring Observatory on 2022 May 08 (proposal ID 2220157, PI Miller-Jones). We obtained 2x2400\,s exposures using the R3000 and B3000 gratings and a NeAr arc lamp exposure immediately following the target exposures. The data were reduced using standard procedures including the PyWiFeS reduction pipeline \citep{childress_pywifes_2014}. LTT4364 was used as the flux standard and a quartz-iodine lamp was used for flat-fielding. We then chose the slitlets with the most significant flux from the calibrated spectra obtained from the pipeline and performed background subtraction, resulting in a spectrum with spectral range 3500 to 9000\,\AA.

\begin{figure}
    \centering
    \includegraphics[scale=0.8]{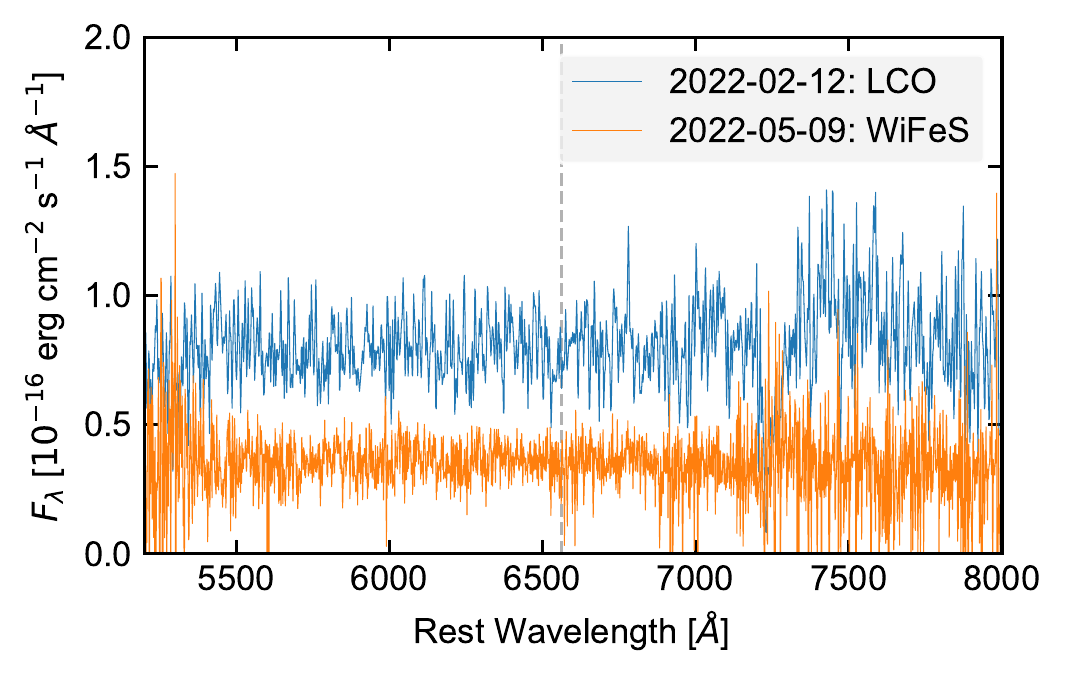}
    \caption{Optical spectra of J1331, with the first follow-up spectrum being obtained on 2022-02-12, $\sim$23 days after the last eRASS5 detection.}
    \label{fig:optical_spectra}
\end{figure}

Each follow-up optical spectrum appears to be consistent with a quiescent host galaxy (Fig.~\ref{fig:optical_spectra}), with no TDE-like optical emission features detected, nor any transient features relative to the NOT spectrum taken on 1999-01-26 and presented in \citet{hampel_new_2022}.

\section{Inferring the outburst properties}\label{sec:outburst_inference}
\begin{figure}
    \centering
    \includegraphics[scale=0.8]{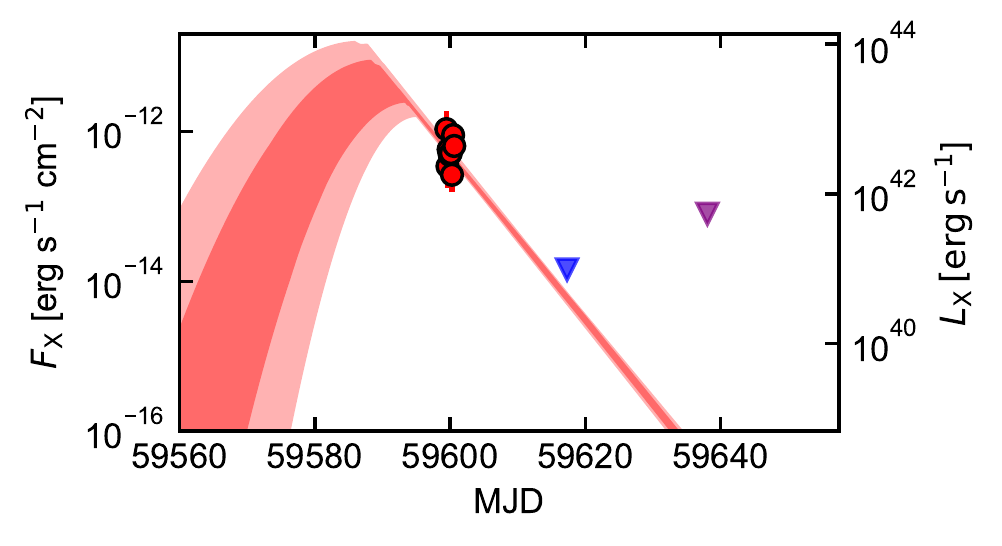}
    \caption{Inferred full outburst (red) for the flaring observed by eROSITA in 2022, assuming the model described in equation~\ref{equation:outburst}. The markers follow the same legend as for Fig.~\ref{fig:long_term_lightcurve}. The darker and lighter shaded red bands enclose the inner 68\% and 98\% of the posterior.}
    \label{fig:reconstructed_outburst}
\end{figure}

To obtain a coarse reconstruction of the 2022 outburst, we perform a joint fit of the rising lightcurve from 1993, observed by \textit{ROSAT}, and the decay lightcurve from 2022, observed by eROSITA and \textit{XMM}, using:
\begin{equation}\label{equation:outburst}
    F_{\mathrm{X}}(t)=F_{\mathrm{X, max}} \times \begin{cases}
\exp \left[ -(t-t_{\mathrm{peak,1}})^2/2 \sigma^2 \right] & \mathrm{if}~t<t_{\mathrm{peak,1}} \\
\exp \left[ -(t-t_{\mathrm{peak,2}})/ \tau \right] & \mathrm{if}~t>t_{\mathrm{peak,2}}
\end{cases}
\end{equation}

where the free parameters of this model are $\sigma$ (the rise timescale), $t_{\mathrm{peak,1}}$ and $t_{\mathrm{peak,2}}$ (the peak time of the \textit{ROSAT} and eROSITA outbursts, respectively), $\tau$ (the decay timescale), and $F_{\mathrm{X, max}}$ (the peak flux of both outbursts), with the priors on these parameters listed in Table~\ref{tab:lightcurve_priors}. We assume that the upper bound on the peak luminosity must be less than the Eddington luminosity for the SMBH, and that both outbursts have the same peak luminosity. We then assume that the rise for 2022 outburst was similar to the 1993 outburst (see below), and use its modelled rise to approximate that of the \textit{unobserved} rise of the 2022 outburst. From this fitted lightcurve model (Fig.~\ref{fig:reconstructed_outburst}), we then computed the integrated 0.2--2~keV luminosity, and corrected this to a bolometric luminosity using the best fitting X-ray spectral model. The inferred energy emitted in each outburst is $(5^{+6}_{-3})\times 10^{49}$~erg, corresponding to an accreted mass of $(5^{+7}_{-2})\times 10^{-4} ( \epsilon / 0.05)^{-1}$~ $\mathrm{M}_{\odot}$, where $\epsilon$ is the radiative efficiency of accretion, whilst the inferred peak MJD for each outburst are $49024 ^{+6}_{-6}$ and $59593 ^{+3}_{-2}$. The inferred $\sigma$ and $\tau$ are $6 ^{+1}_{-1}$~days and $3.9^{+0.1}_{-0.1}$~days, respectively, and we roughly estimate the MJD of disruption to be $59593 - 2*\sigma \sim 59581$.
\begin{table}
        \centering
        \caption{Priors adopted in the fitting of the 1993 and 2022 outbursts. The rise and decay timescales are in units of days. $t_{\mathrm{peak,1}}$ and $t_{\mathrm{peak,2}}$ are in MJD, whilst $F_{\mathrm{max}}$ is the maximum 0.2--2~keV flux of each outburst (with upper bound set by the Eddington luminosity of the system).}  
        \label{tab:lightcurve_priors}
        \begin{tabular}{lccr} 
                \hline
                Parameter  & Prior \\
                \hline
                $\log[ \sigma ]$ & \hbox{\strut $\sim \mathcal{U}(0, \log [50])$} \\
                $t_{\mathrm{peak,1}}$ & \hbox{\strut $\sim \mathcal{U}(49006, 49178)$} \\
                $t_{\mathrm{peak,2}}$ & \hbox{\strut $\sim \mathcal{U}(58450, 58650)$} \\
                $\log[ \tau ]$ & \hbox{\strut $\sim \mathcal{U}(0, \log [50])$}  \\
                $\log[ F_{\mathrm{X, max}} ]$ & 
                \hbox{\strut $\sim \mathcal{U}(\log [5\times10^{-13}],\log [4\times10^{-11}])$} \\
                \hline
        \end{tabular}
\end{table}

It is of course extremely important to consider that these estimates are subject to a number of caveats, mainly related to our observations not covering the rise of the 2022 outburst, such that the estimated values here should be treated with caution. For example, it is assumed that the outburst can be well modelled by equation~\ref{equation:outburst}, and that both the 1993 and 2022 outbursts are similar, whereas the actual lightcurve may have had an extended plateau phase prior to the eROSITA detection (so our estimated fluence and accreted mass would be underestimated). 

However, if the 2022 outburst does evolve relatively closely to the functional form in equation~\ref{equation:outburst}, then it may be reasonable to consider that the rise timescale for the flare in 1993 is similar to that observed in 2022 (under a tidal disruption scenario), due to the approximately constant eccentricity of the stellar remnant after repeated partial disruptions \citep{antonini_tidal_2011}, and the weak dependence of the period of the most bound debris on the stellar mass \citep{hayasaki_finite_2013}.

\begin{figure}
    \centering
    \includegraphics[scale=0.999]{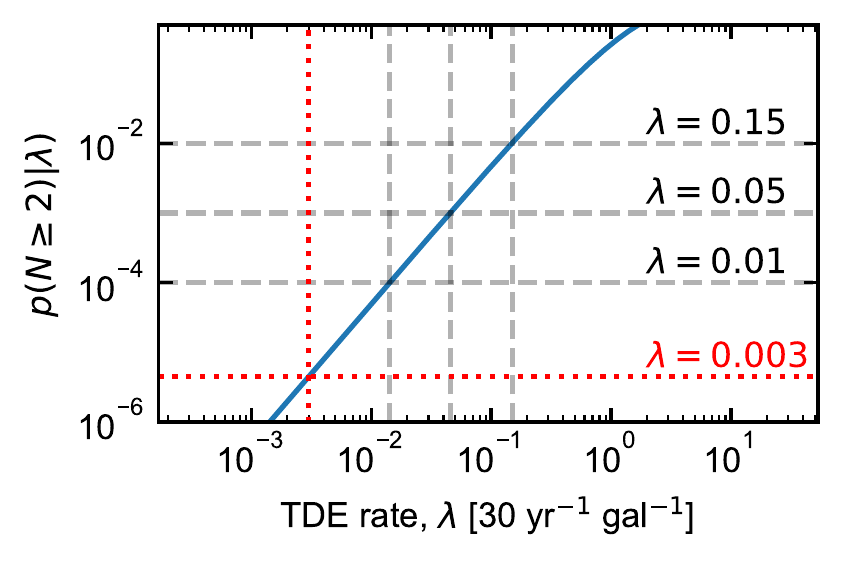}
    \caption{Poisson probability of $N\geq 2$ TDEs occurring within a 30 year period for a given galaxy. The red dotted lines mark the estimated probability for current theoretical estimates for TDE rates ($10^{-4}$~yr$^{-1}$~gal$^{-1}$; \citealt{stone_rates_2020}). The grey dashed lines mark out the TDE rates of 0.15, 0.05 and 0.01 per ~30~yr$^{-1}$~gal$^{-1}$, required to produce probabilities of 0.01, 0.001, and 0.0001, respectively.}
    \label{fig:poisson_probability_tde_rate}
\end{figure}

\section{Additional X-ray information}
\begin{figure}
    \centering
    \includegraphics{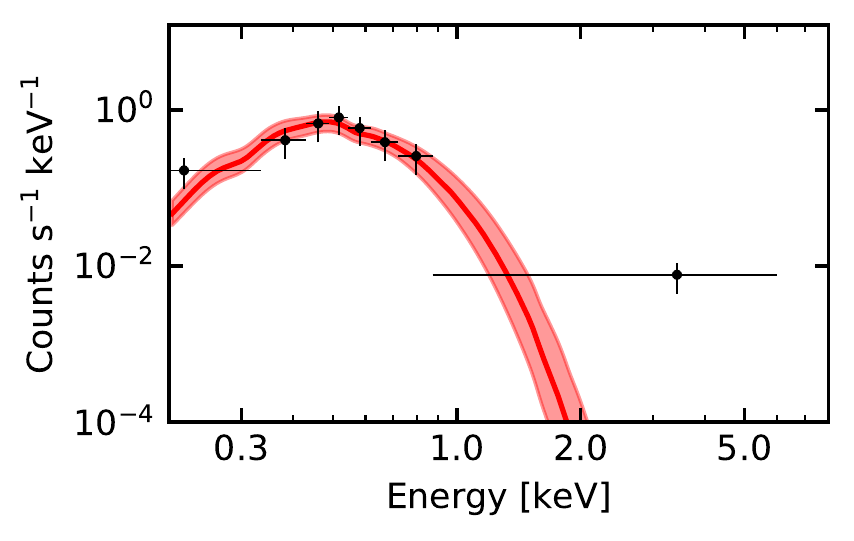}
    \caption{BXA fit of a \texttt{tbabs*zbbody} model to the eRASS5 spectrum. The solid red line represents the median model fit, whilst the shaded red region encloses the inner 98\% of the credible region. The X-ray spectrum is ultra-soft with $kT=0.115 ^{+0.007}_{-0.007} $~keV.}
    \label{fig:bxa_fit_erass5}
\end{figure}

\begin{figure}
    \centering
    \includegraphics[scale=0.8]{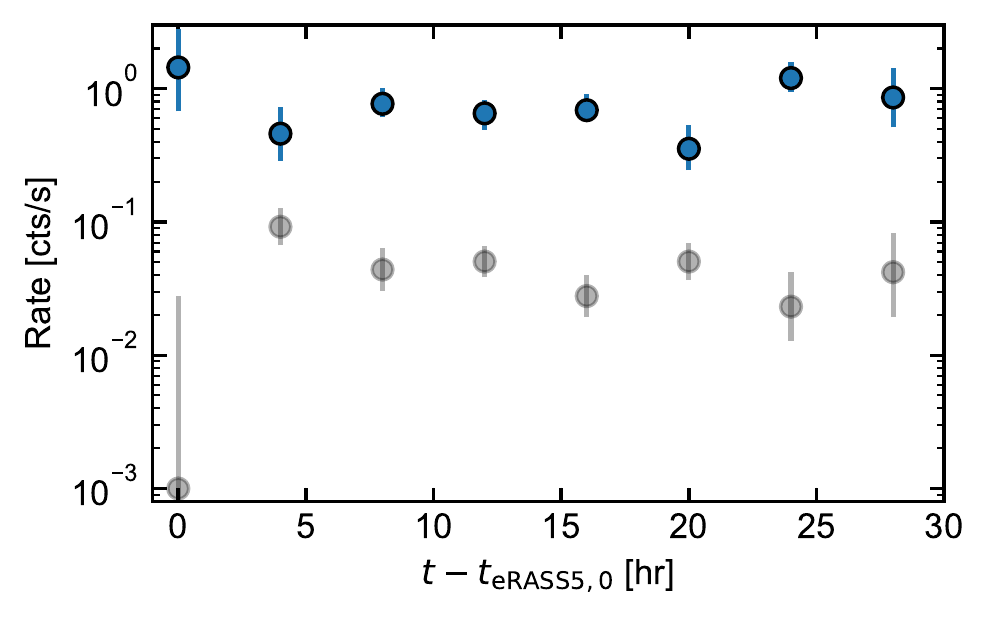}
    \caption{0.2--2~keV band eRASS5 lightcurve of J1331. The blue and grey markers denote the inferred source and background count rates in the source aperture, respectively. Times are measured relative to the start of the earliest observation of J1331 in eRASS5, $t_{\mathrm{eRASS5,0}}$. J1331 is clearly detected above background in each visit.}
    \label{fig:erass5_lightcurve}
\end{figure}

\begin{figure}
    \centering
    \includegraphics[scale=0.8]{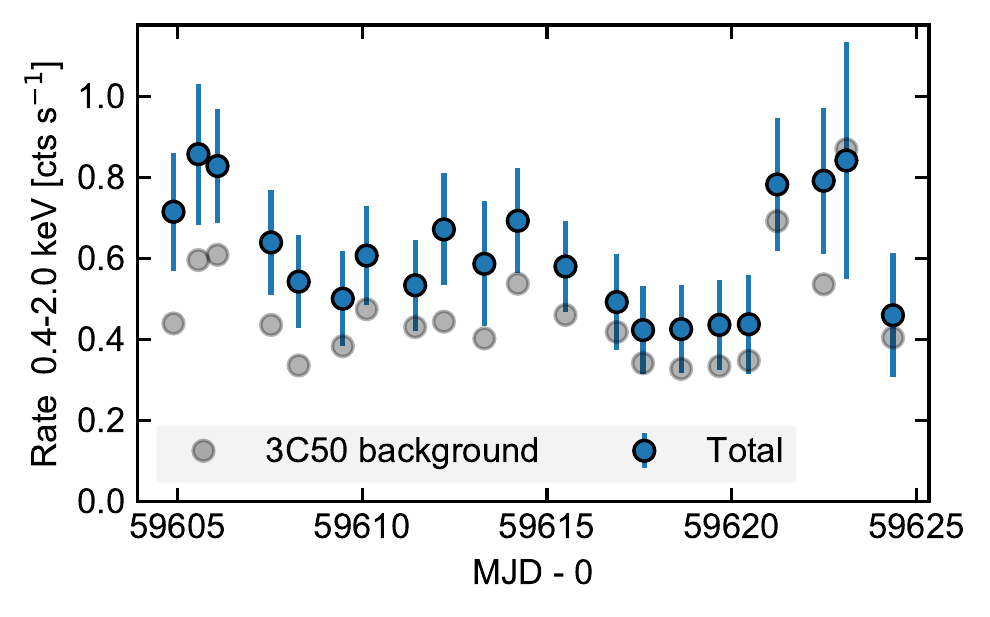}
    \caption{\textit{NICER} count rate lightcurve in the 0.4-2~keV band, with blue markers denoting the total observed count rate (source and background), and grey markers representing the estimated background rate inferred using the 3C50 background model \citep{remillard_empirical_2022}. The system is not detected at 2$\sigma$ above background in each \textit{NICER} OBSID.}
    \label{fig:nicer_rate_lc}
\end{figure}

\begin{figure}
    \centering
    \includegraphics[scale=1]{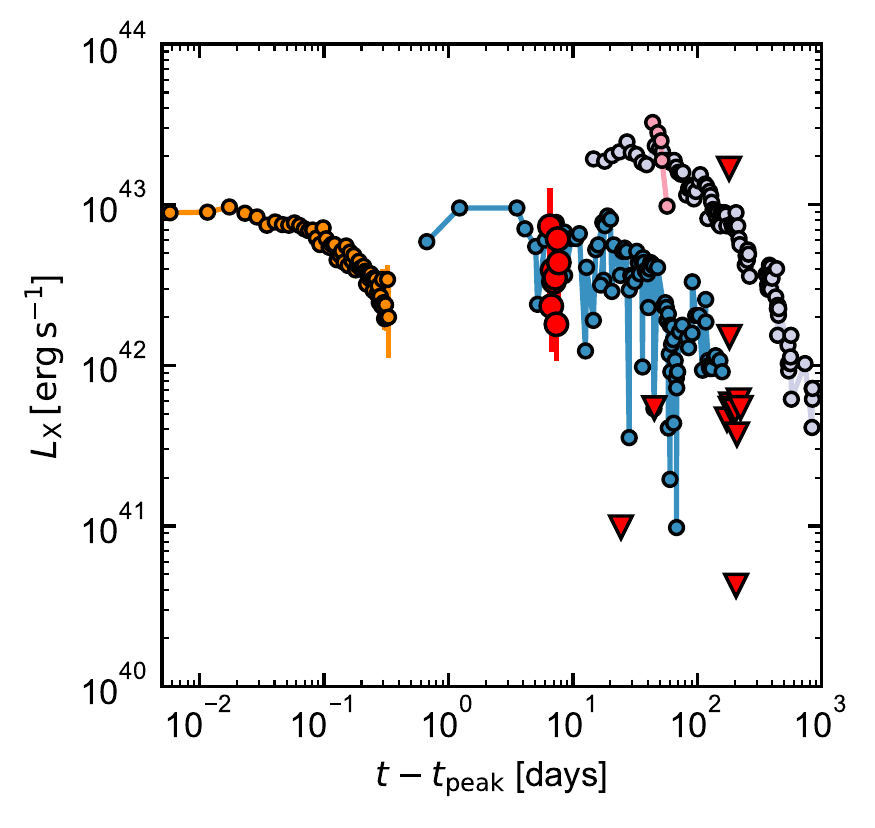}
    \caption{Comparison of the 0.2--2~keV X-ray lightcurve evolution of J1331 (red markers) with other soft nuclear transients from quiescent galaxies (or those recently hosting low luminosity AGN). J1331 decays in $L_{\mathrm{X}}$ over longer timescales than QPEs (orange for eROQPE1; \citealt{arcodia_x-ray_2021}), but still over much shorter timescales than previously reported TDEs in the literature, such as ASAS-SN~14li (grey, \citealt{bright_long-term_2018}), AT~2019azh decay phase (blue, \citealt{hinkle_discovery_2020}), AT~2019dsg (pink, \citealt{cannizzaro_accretion_2021}). The $t_{\mathrm{peak}}$ for J1331 was set to MJD=59592.9, following the assumptions described in Section~\ref{sec:outburst_inference}.}
    \label{fig:nuclear_transient_comparison}
\end{figure}

The BXA fitted model to the eRASS5 spectrum is shown in Fig.~\ref{fig:bxa_fit_erass5}, and the eRASS5 lightcurve is shown in Fig.~\ref{fig:erass5_lightcurve}. The \textit{NICER} count rate lightcurve is plotted in Fig.~\ref{fig:nicer_rate_lc}, whilst the full X-ray lightcurve of J1331 is presented in Table~\ref{tab:x_ray_lightcurve_table}. A comparison of the X-ray lightcurve of J1331 with other nuclear transients is presented in Fig~\ref{fig:nuclear_transient_comparison}.

\begin{table}
\centering
\caption{X-ray lightcurve table for J1331. The fluxes from the \textit{ROSAT} pointed observations were derived from \citet{hampel_new_2022}. The first four eROSITA observations listed, between MJD 58868 and 59419, are upper limits estimated from eRASS1, 2, 3 and 4, respectively; eROSITA fluxes outside of this window have been computed from the individual visits within eRASS5.}
\label{tab:x_ray_lightcurve_table}
\begin{tabular}{cccc}
\hline
MJD & Observation & $F_{\mathrm{0.2-2keV, obs}}$ & $F_{\mathrm{0.2-2keV, unabs}}$ \\
 & & [$10^{-13}$\,erg cm$^{-2}$ s$^{-1}$] & [$10^{-13}$\,erg cm$^{-2}$ s$^{-1}$] \\
\hline
48260.000 & \textit{ROSAT}/ RASS & $< 2.9$ & $< 4.5 $ \\
48844.598 & \textit{ROSAT}/ Pointed & $< 0.2$ & $< 0.4 $ \\
49006.094 & \textit{ROSAT}/ Pointed & $< 1.2$ & $< 1.9 $ \\
49012.146 & \textit{ROSAT}/ Pointed & $6.1 \pm 0.7$ & $9.4 \pm 1.0$ \\
49012.180 & \textit{ROSAT}/ Pointed & $8.9 \pm 1.9$ & $13.8 \pm 2.9$ \\
49013.591 & \textit{ROSAT}/ Pointed & $10.0 \pm 1.1$ & $15.5 \pm 1.7$ \\
49178.555 & \textit{ROSAT}/ Pointed & $< 0.7$ & $< 1.0 $ \\
49178.766 & \textit{ROSAT}/ Pointed & $< 0.3$ & $< 0.5 $ \\
53745.291 & \textit{XMM}/ Slew & $< 3.8$ & $< 5.9 $ \\
57056.039 & \textit{XMM}/ Slew & $< 5.4$ & $< 8.3 $ \\
57241.869 & \textit{XMM}/ Slew & $< 8.3$ & $< 12.8 $ \\
58226.719 & \textit{Swift}/ XRT & $< 0.9$ & $< 1.4 $ \\
58230.707 & \textit{Swift}/ XRT & $< 0.5$ & $< 0.8 $ \\
58234.028 & \textit{Swift}/ XRT & $< 0.8$ & $< 1.2 $ \\
58868.114 & \textit{SRG}/ eROSITA & $<0.3$ & $<0.4$ \\
59051.625 & \textit{SRG}/ eROSITA & $<0.5$ & $<0.7$ \\
59229.875 & \textit{SRG}/ eROSITA & $<1.3$ & $<1.7$  \\
59418.532 & \textit{SRG}/ eROSITA & $<0.5$ & $<0.7$  \\
59599.448 & \textit{SRG}/ eROSITA & $10.8 \pm 8.0$ & $14.4 \pm 10.6$ \\
59599.614 & \textit{SRG}/ eROSITA & $3.4 \pm 1.7$ & $4.6 \pm 2.2$ \\
59599.781 & \textit{SRG}/ eROSITA & $5.7 \pm 1.5$ & $7.7 \pm 2.0$ \\
59599.948 & \textit{SRG}/ eROSITA & $4.9 \pm 1.2$ & $6.5 \pm 1.6$ \\
59600.114 & \textit{SRG}/ eROSITA & $5.1 \pm 1.3$ & $6.9 \pm 1.7$ \\
59600.281 & \textit{SRG}/ eROSITA & $2.6 \pm 1.1$ & $3.5 \pm 1.5$ \\
59600.448 & \textit{SRG}/ eROSITA & $9.0 \pm 2.4$ & $12.0 \pm 3.2$ \\
59600.614 & \textit{SRG}/ eROSITA & $6.4 \pm 3.5$ & $8.5 \pm 4.6$ \\
59604.892 & \textit{NICER}/ XTI & $<$8.6 & $<$13.8 \\
59605.566 & \textit{NICER}/ XTI & $<$10.3 & $<$16.6 \\
59606.082 & \textit{NICER}/ XTI & $<$9.5 & $<$15.3 \\
59607.533 & \textit{NICER}/ XTI & $<$7.7 & $<$12.3 \\
59608.280 & \textit{NICER}/ XTI & $<$6.6 & $<$10.6 \\
59609.473 & \textit{NICER}/ XTI & $<$6.3 & $<$10.1 \\
59610.119 & \textit{NICER}/ XTI & $<$7.3 & $<$11.7 \\
59611.432 & \textit{NICER}/ XTI & $<$6.5 & $<$10.4 \\
59612.210 & \textit{NICER}/ XTI & $<$8.1 & $<$13.0 \\
59613.305 & \textit{NICER}/ XTI & $<$7.6 & $<$12.3 \\
59614.210 & \textit{NICER}/ XTI & $<$8.1 & $<$13.1 \\
59615.500 & \textit{NICER}/ XTI & $<$6.9 & $<$11.1 \\
59616.889 & \textit{NICER}/ XTI & $<$6.2 & $<$10.0 \\
59617.287 & \textit{XMM}/ Pointed & $< 0.1$ & $< 0.2 $ \\
59617.598 & \textit{NICER}/ XTI & $<$5.5 & $<$8.8 \\
59618.630 & \textit{NICER}/ XTI & $<$5.5 & $<$8.8 \\
59619.666 & \textit{NICER}/ XTI & $<$5.6 & $<$9.0 \\
59620.463 & \textit{NICER}/ XTI & $<$5.8 & $<$9.4 \\
59621.229 & \textit{NICER}/ XTI & $<$9.5 & $<$15.3 \\
59622.488 & \textit{NICER}/ XTI & $<$9.8 & $<$15.8 \\
59623.102 & \textit{NICER}/ XTI & $<$12.2 & $<$19.6 \\
59624.362 & \textit{NICER}/ XTI & $<$6.5 & $<$10.5 \\
59638.031 & \textit{Swift}/ XRT & $< 0.8$ & $< 1.4 $ \\
59766.375 & \textit{Swift}/ XRT & $< 0.7$ & $< 1.2 $ \\
59773.061 & \textit{Swift}/ XRT & $< 24.6$ & $< 43.7 $ \\
59774.292 & \textit{Swift}/ XRT & $< 2.2$ & $< 3.9 $ \\
59778.974 & \textit{Swift}/ XRT & $< 0.8$ & $< 1.4 $ \\
59780.760 & \textit{Swift}/ XRT & $< 0.8$ & $< 1.5 $ \\
59787.468 & \textit{Swift}/ XRT & $< 0.8$ & $< 1.4 $ \\
59794.352 & \textit{Swift}/ XRT & $< 0.8$ & $< 1.4 $ \\
59797.916 & \textit{XMM}/ Pointed & $< 0.06$ & $< 0.10 $ \\
59801.282 & \textit{Swift}/ XRT & $< 0.5$ & $< 1.0 $ \\
59808.180 & \textit{Swift}/ XRT & $< 0.9$ & $< 1.6 $ \\
59815.534 & \textit{Swift}/ XRT & $< 0.8$ & $< 1.4 $ \\
\end{tabular}
\end{table}

\section{Additional photometric information}
\begin{figure}
    \centering
    \includegraphics[scale=0.7]{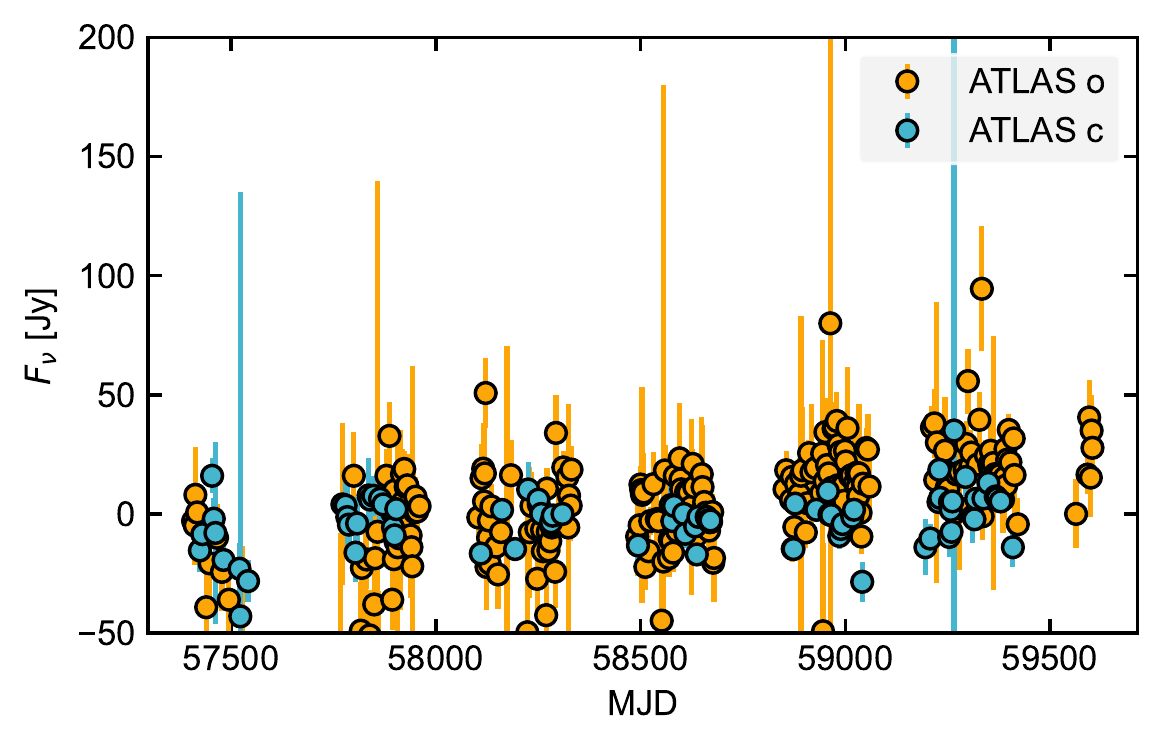}
    \includegraphics[scale=0.8]{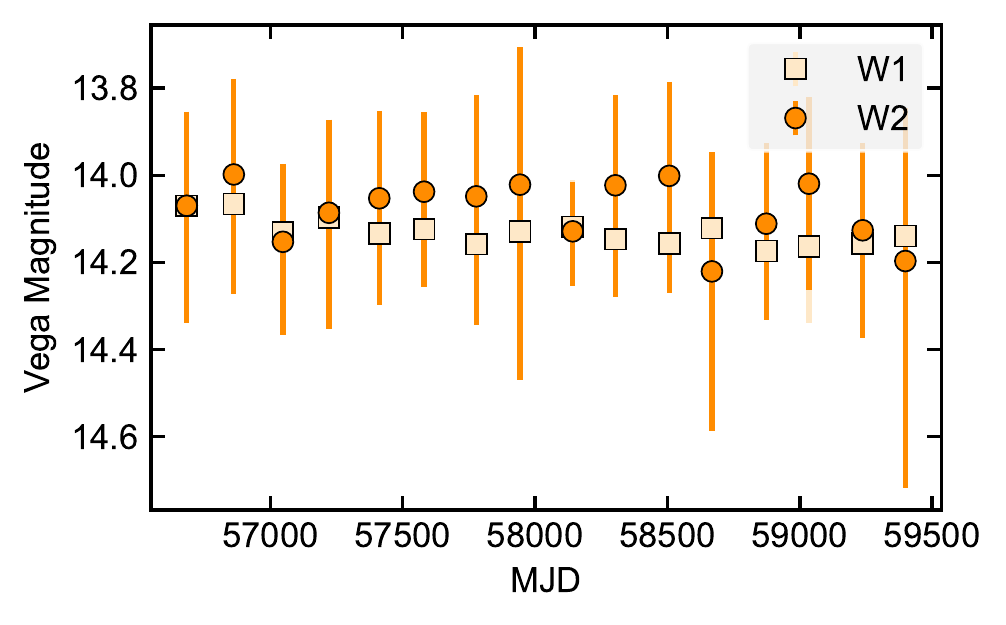}
    \caption{No major variability is seen within the ATLAS forced photometry generated on the difference imaging (top), nor within the \textit{NEOWISE} lightcurve (bottom).}
    \label{fig:atlas_neowise_lightcurve}
\end{figure}

Table~\ref{tab:uvot_photometry} contains the \textit{Swift} UVOT aperture photometry of the host galaxy of J1331, whilst Fig.~\ref{fig:atlas_neowise_lightcurve} shows the long term ATLAS and \textit{NEOWISE} lightcurves of J1331.
\begin{table}
\centering
\caption{\textit{Swift} UVM2 photometry of the host galaxy of J1331.}
\label{tab:uvot_photometry}
\begin{tabular}{cc}
\hline
MJD & Magnitude \\
\hline
58226.727 & 23.1 $\pm$1.0 \\
58230.747 & 22.9 $\pm$0.6 \\
58234.068 & 22.3 $\pm$0.4 \\
59638.032 & 22.2 $\pm$0.3 \\
59766.376 & 23.0 $\pm$0.7 \\
59774.294 & 22.8 $\pm$1.0 \\
59778.975 & 22.5 $\pm$0.6 \\
59780.762 & 22.9 $\pm$0.8 \\
59794.353 & 22.7 $\pm$0.6 \\
59801.283 & 22.5 $\pm$0.4 \\
59808.181 & 22.8 $\pm$0.6 \\
59815.535 & 22.5 $\pm$0.5 \\
\end{tabular}
\end{table}

\bsp	
\label{lastpage}
\end{document}